\documentstyle[10pt,epsfig,dp_delphititle,float,hangcaption,xspace,amssymb,
amsfonts,amsmath,amsthm,cite,graphicx,lineno]{dp_delphi}
\makeindex
\def\DpPaperGroup{EP}
\def\DpPaperRef{2003-044}
\def\DpDate{21 Avril 2003}
\def\DpAuthors{DELPHI Collaboration}
\def\DpSubmit{(Accepted by Phys. Lett. B)}
\def\DpTitle{{A measurement of the branching fractions of the $b$-quark 
       into charged and neutral $b$-hadrons}}

\newcommand{\degr}{$^\circ$\xspace}
\newcommand{\jst}{\textsc{Jetset 7.3}\xspace}

\newcommand{\delphi}{\textsc{DELPHI}\xspace}
\newcommand{\lthree}{\textsc{L3}\xspace}
\newcommand{\cdf}{\textsc{CDF}\xspace}
\newcommand{\sld}{\textsc{SLD}\xspace}
\newcommand{\opal}{\textsc{OPAL}\xspace}
\newcommand{\alephlep}{\textsc{ALEPH}\xspace}
\newcommand{\delsim}{\textsc{Delsim}\xspace}

\newcommand{\cc}{{{c}\overline{{c}}}\xspace}
\newcommand{\bb}{{{b}\overline{{b}}}\xspace}
\newcommand{\sq}{{{s}\overline{{s}}}\xspace}
\newcommand{\uu}{{{u}\overline{{u}}}\xspace}
\newcommand{\dd}{{{d}\overline{{d}}}\xspace}

\newcommand{\bbar}{{\overline{{b}}}\xspace}

\newcommand{\bplus}{{{B}^+}\xspace}

\newcommand{\bnull}{{{B}^0}\xspace}
\newcommand{\bnullbar}{{\overline{{B}}^0}\xspace}

\newcommand{\bnulls}{{{B}^0_s}\xspace}
\newcommand{\bnullsbar}{{\overline{{B}}^0_s}\xspace}

\newcommand{\dbarnull}{{\overline{{D}}^{0}}\xspace}
\newcommand{\dminus}{{{D}^{-}}\xspace}

\newcommand{\bdstarud}{{{B}^{**}_{u,d}}\xspace}

\newcommand{\lambdab}{{\Lambda_b}\xspace}

\newcommand{\fbs}{{f_{B_s}}\xspace}
\newcommand{\fbsprime}{{f'_{B_s}}\xspace}
\newcommand{\fbu}{{f_{B_u}}\xspace}
\newcommand{\fbuprime}{{f'_{B_u}}\xspace}
\newcommand{\fbd}{{f_{B_d}}\xspace}
\newcommand{\fbdprime}{{f'_{B_d}}\xspace}

\newcommand{\fbb}{{f_{b-baryon}}\xspace}
\newcommand{\fbbnull}{{f_{b-baryon}^0}\xspace}
\newcommand{\fbbplus}{{f_{b-baryon}^+}\xspace}

\newcommand{\fnull}{{f^0}\xspace}
\newcommand{\fplus}{{f^+}\xspace}
\newcommand{\fbiprime}{{f'_{i}}\xspace}
\newcommand{\fbi}{{f_{i}}\xspace}
\newcommand{\NIM}[3]{{\em Nucl. Instrum. Methods} {\bf{#1} }{(#2) }{#3}}

\newcommand{\PLB}[3]{{\em Phys. Lett.} {\bf B}{\bf{#1} }{(#2) }{#3}}
\newcommand{\PRL}[3]{{\em Phys. Rev. Lett.} {\bf{#1} }{(#2) }{#3}}
\newcommand{\PRD}[3]{{\em Phys. Rev.} {\bf D}{\bf{#1} }{(#2) }{#3}}
\newcommand{\ZPC}[3]{{\em Z. Phys.} {\bf C}{\bf{#1} }{(#2) }{#3}}

\newcommand{\CPC}[3]{{\em Comp. Phys. Comm.} {\bf{#1} }{(#2) }{#3}}
\newcommand{\EJC}[3]{{\em Eur. Phys. J.} {\bf C}{\bf{#1} }{(#2) }{#3}}
\newcommand{\EJCNP}{{\em Eur. Phys. J.} {\bf C}}
\begin{document}
\makeatletter
\newcount\@tempcntc
\def\@citex[#1]#2{\if@filesw\immediate\write\@auxout{\string\citation{#2}}\fi
  \@tempcnta\z@\@tempcntb\m@ne\def\@citea{}\@cite{\@for\@citeb:=#2\do
    {\@ifundefined
       {b@\@citeb}{\@citeo\@tempcntb\m@ne\@citea\def\@citea{,}{\bf ?}\@warning
       {Citation `\@citeb' on page \thepage \space undefined}}%
    {\setbox\z@\hbox{\global\@tempcntc0\csname b@\@citeb\endcsname\relax}%
     \ifnum\@tempcntc=\z@ \@citeo\@tempcntb\m@ne
       \@citea\def\@citea{,}\hbox{\csname b@\@citeb\endcsname}%
     \else
      \advance\@tempcntb\@ne
      \ifnum\@tempcntb=\@tempcntc
      \else\advance\@tempcntb\m@ne\@citeo
      \@tempcnta\@tempcntc\@tempcntb\@tempcntc\fi\fi}}\@citeo}{#1}}
\def\@citeo{\ifnum\@tempcnta>\@tempcntb\else\@citea\def\@citea{,}%
  \ifnum\@tempcnta=\@tempcntb\the\@tempcnta\else
   {\advance\@tempcnta\@ne\ifnum\@tempcnta=\@tempcntb \else \def\@citea{--}\fi
    \advance\@tempcnta\m@ne\the\@tempcnta\@citea\the\@tempcntb}\fi\fi}
 
\makeatother
\begin{titlepage}
\pagenumbering{roman}
\CERNpreprint{\DpPaperGroup}{\DpPaperRef} 
\date{{\small\DpDate}} 
\title{\DpTitle} 
\address{\DpAuthors} 
\begin{shortabs} 
\noindent
%
\noindent

The production fractions of charged and neutral
$b$-hadrons in $b$-quark events from $Z^0$ decays
have been measured with the DELPHI detector at LEP.
An algorithm has been developed, based on a neural network,
to estimate the charge of the weakly-decaying $b$-hadron
by distinguishing its decay products from
particles produced at the primary vertex.
From the data taken in the years 1994 and 1995, the fraction of
$\bbar$-quarks fragmenting into
positively charged
weakly-decaying
$b$-hadrons has been measured to be:
\begin{center}
 $\fplus = (42.09 \pm 0.82 \mbox{(stat.)} \pm 0.89 \mbox{(syst.)})\%$. 
\end{center}
Subtracting the rates for charged
$\overline{\Xi}_b^+$ and $\overline{\Omega}_b^+$ baryons gives the production fraction
of $\bplus$ mesons:
\begin{center}
 $\fbu = (40.99 \pm 0.82 \mbox{(stat.)} \pm 1.11 \mbox{(syst.)})\%$. 
\end{center}
\end{shortabs}
\vfill
\begin{center}
\DpSubmit \ \\ 
\end{center}
\vfill
\clearpage
\headsep 10.0pt
\addtolength{\textheight}{10mm}
\addtolength{\footskip}{-5mm}
\begingroup
%
\newcommand{\DpName}[2]{\hbox{#1$^{\ref{#2}}$},\hfill}
\newcommand{\DpNameTwo}[3]{\hbox{#1$^{\ref{#2},\ref{#3}}$},\hfill}
\newcommand{\DpNameThree}[4]{\hbox{#1$^{\ref{#2},\ref{#3},\ref{#4}}$},\hfill}
\newskip\Bigfill \Bigfill = 0pt plus 1000fill
\newcommand{\DpNameLast}[2]{\hbox{#1$^{\ref{#2}}$}\hspace{\Bigfill}}
%
\footnotesize
\noindent
\DpName{J.Abdallah}{LPNHE}
\DpName{P.Abreu}{LIP}
\DpName{W.Adam}{VIENNA}
\DpName{P.Adzic}{DEMOKRITOS}
\DpName{T.Albrecht}{KARLSRUHE}
\DpName{T.Alderweireld}{AIM}
\DpName{R.Alemany-Fernandez}{CERN}
\DpName{T.Allmendinger}{KARLSRUHE}
\DpName{P.P.Allport}{LIVERPOOL}
\DpName{U.Amaldi}{MILANO2}
% ...
\normalsize
\endgroup
\titlefoot{Department of Physics and Astronomy, Iowa State
     University, Ames IA 50011-3160, USA
    \label{AMES}}
\titlefoot{Physics Department, Universiteit Antwerpen,
     Universiteitsplein 1, B-2610 Antwerpen, Belgium \\
     \indent~~and IIHE, ULB-VUB,
     Pleinlaan 2, B-1050 Brussels, Belgium \\
     \indent~~and Facult\'e des Sciences,
     Univ. de l'Etat Mons, Av. Maistriau 19, B-7000 Mons, Belgium
    \label{AIM}}
\titlefoot{Physics Laboratory, University of Athens, Solonos Str.
     104, GR-10680 Athens, Greece
    \label{ATHENS}}
\titlefoot{Department of Physics, University of Bergen,
     All\'egaten 55, NO-5007 Bergen, Norway
    \label{BERGEN}}
\titlefoot{Dipartimento di Fisica, Universit\`a di Bologna and INFN,
     Via Irnerio 46, IT-40126 Bologna, Italy
    \label{BOLOGNA}}
\titlefoot{Centro Brasileiro de Pesquisas F\'{\i}sicas, rua Xavier Sigaud 150,
     BR-22290 Rio de Janeiro, Brazil \\
     \indent~~and Depto. de F\'{\i}sica, Pont. Univ. Cat\'olica,
     C.P. 38071 BR-22453 Rio de Janeiro, Brazil \\
     \indent~~and Inst. de F\'{\i}sica, Univ. Estadual do Rio de Janeiro,
     rua S\~{a}o Francisco Xavier 524, Rio de Janeiro, Brazil
    \label{BRASIL}}
\titlefoot{Coll\`ege de France, Lab. de Physique Corpusculaire, IN2P3-CNRS,
     FR-75231 Paris Cedex 05, France
    \label{CDF}}
\titlefoot{CERN, CH-1211 Geneva 23, Switzerland
    \label{CERN}}
\titlefoot{Institut de Recherches Subatomiques, IN2P3 - CNRS/ULP - BP20,
     FR-67037 Strasbourg Cedex, France
    \label{CRN}}
\titlefoot{Now at DESY-Zeuthen, Platanenallee 6, D-15735 Zeuthen, Germany
    \label{DESY}}
\titlefoot{Institute of Nuclear Physics, N.C.S.R. Demokritos,
     P.O. Box 60228, GR-15310 Athens, Greece
    \label{DEMOKRITOS}}
\titlefoot{FZU, Inst. of Phys. of the C.A.S. High Energy Physics Division,
     Na Slovance 2, CZ-180 40, Praha 8, Czech Republic
    \label{FZU}}
\titlefoot{Dipartimento di Fisica, Universit\`a di Genova and INFN,
     Via Dodecaneso 33, IT-16146 Genova, Italy
    \label{GENOVA}}
\titlefoot{Institut des Sciences Nucl\'eaires, IN2P3-CNRS, Universit\'e
     de Grenoble 1, FR-38026 Grenoble Cedex, France
    \label{GRENOBLE}}
\titlefoot{Helsinki Institute of Physics, P.O. Box 64,
     FIN-00014 University of Helsinki, Finland
    \label{HELSINKI}}
\titlefoot{Joint Institute for Nuclear Research, Dubna, Head Post
     Office, P.O. Box 79, RU-101 000 Moscow, Russian Federation
    \label{JINR}}
\titlefoot{Institut f\"ur Experimentelle Kernphysik,
     Universit\"at Karlsruhe, Postfach 6980, DE-76128 Karlsruhe,
     Germany
    \label{KARLSRUHE}}
\titlefoot{Institute of Nuclear Physics,Ul. Kawiory 26a,
     PL-30055 Krakow, Poland
    \label{KRAKOW1}}
\titlefoot{Faculty of Physics and Nuclear Techniques, University of Mining
     and Metallurgy, PL-30055 Krakow, Poland
    \label{KRAKOW2}}
\titlefoot{Universit\'e de Paris-Sud, Lab. de l'Acc\'el\'erateur
     Lin\'eaire, IN2P3-CNRS, B\^{a}t. 200, FR-91405 Orsay Cedex, France
    \label{LAL}}
\titlefoot{School of Physics and Chemistry, University of Lancaster,
     Lancaster LA1 4YB, UK
    \label{LANCASTER}}
\titlefoot{LIP, IST, FCUL - Av. Elias Garcia, 14-$1^{o}$,
     PT-1000 Lisboa Codex, Portugal
    \label{LIP}}
\titlefoot{Department of Physics, University of Liverpool, P.O.
     Box 147, Liverpool L69 3BX, UK
    \label{LIVERPOOL}}
\titlefoot{Dept. of Physics and Astronomy, Kelvin Building,
     University of Glasgow, Glasgow G12 8QQ
    \label{GLASGOW}}
\titlefoot{LPNHE, IN2P3-CNRS, Univ.~Paris VI et VII, Tour 33 (RdC),
     4 place Jussieu, FR-75252 Paris Cedex 05, France
    \label{LPNHE}}
\titlefoot{Department of Physics, University of Lund,
     S\"olvegatan 14, SE-223 63 Lund, Sweden
    \label{LUND}}
\titlefoot{Universit\'e Claude Bernard de Lyon, IPNL, IN2P3-CNRS,
     FR-69622 Villeurbanne Cedex, France
    \label{LYON}}
\titlefoot{Dipartimento di Fisica, Universit\`a di Milano and INFN-MILANO,
     Via Celoria 16, IT-20133 Milan, Italy
    \label{MILANO}}
\titlefoot{Dipartimento di Fisica, Univ. di Milano-Bicocca and
     INFN-MILANO, Piazza della Scienza 2, IT-20126 Milan, Italy
    \label{MILANO2}}
\titlefoot{IPNP of MFF, Charles Univ., Areal MFF,
     V Holesovickach 2, CZ-180 00, Praha 8, Czech Republic
    \label{NC}}
\titlefoot{NIKHEF, Postbus 41882, NL-1009 DB
     Amsterdam, The Netherlands
    \label{NIKHEF}}
\titlefoot{National Technical University, Physics Department,
     Zografou Campus, GR-15773 Athens, Greece
    \label{NTU-ATHENS}}
\titlefoot{Physics Department, University of Oslo, Blindern,
     NO-0316 Oslo, Norway
    \label{OSLO}}
\titlefoot{Dpto. Fisica, Univ. Oviedo, Avda. Calvo Sotelo
     s/n, ES-33007 Oviedo, Spain
    \label{OVIEDO}}
\titlefoot{Department of Physics, University of Oxford,
     Keble Road, Oxford OX1 3RH, UK
    \label{OXFORD}}
\titlefoot{Dipartimento di Fisica, Universit\`a di Padova and
     INFN, Via Marzolo 8, IT-35131 Padua, Italy
    \label{PADOVA}}
\titlefoot{Rutherford Appleton Laboratory, Chilton, Didcot
     OX11 OQX, UK
    \label{RAL}}
\titlefoot{Dipartimento di Fisica, Universit\`a di Roma II and
     INFN, Tor Vergata, IT-00173 Rome, Italy
    \label{ROMA2}}
\titlefoot{Dipartimento di Fisica, Universit\`a di Roma III and
     INFN, Via della Vasca Navale 84, IT-00146 Rome, Italy
    \label{ROMA3}}
\titlefoot{DAPNIA/Service de Physique des Particules,
     CEA-Saclay, FR-91191 Gif-sur-Yvette Cedex, France
    \label{SACLAY}}
\titlefoot{Instituto de Fisica de Cantabria (CSIC-UC), Avda.
     los Castros s/n, ES-39006 Santander, Spain
    \label{SANTANDER}}
\titlefoot{Inst. for High Energy Physics, Serpukov
     P.O. Box 35, Protvino, (Moscow Region), Russian Federation
    \label{SERPUKHOV}}
\titlefoot{J. Stefan Institute, Jamova 39, SI-1000 Ljubljana, Slovenia
     and Laboratory for Astroparticle Physics,\\
     \indent~~Nova Gorica Polytechnic, Kostanjeviska 16a, SI-5000 Nova Gorica, Slovenia, \\
     \indent~~and Department of Physics, University of Ljubljana,
     SI-1000 Ljubljana, Slovenia
    \label{SLOVENIJA}}
\titlefoot{Fysikum, Stockholm University,
     Box 6730, SE-113 85 Stockholm, Sweden
    \label{STOCKHOLM}}
\titlefoot{Dipartimento di Fisica Sperimentale, Universit\`a di
     Torino and INFN, Via P. Giuria 1, IT-10125 Turin, Italy
    \label{TORINO}}
\titlefoot{INFN,Sezione di Torino, and Dipartimento di Fisica Teorica,
     Universit\`a di Torino, Via P. Giuria 1,\\
     \indent~~IT-10125 Turin, Italy
    \label{TORINOTH}}
\titlefoot{Dipartimento di Fisica, Universit\`a di Trieste and
     INFN, Via A. Valerio 2, IT-34127 Trieste, Italy \\
     \indent~~and Istituto di Fisica, Universit\`a di Udine,
     IT-33100 Udine, Italy
    \label{TU}}
\titlefoot{Univ. Federal do Rio de Janeiro, C.P. 68528
     Cidade Univ., Ilha do Fund\~ao
     BR-21945-970 Rio de Janeiro, Brazil
    \label{UFRJ}}
\titlefoot{Department of Radiation Sciences, University of
     Uppsala, P.O. Box 535, SE-751 21 Uppsala, Sweden
    \label{UPPSALA}}
\titlefoot{IFIC, Valencia-CSIC, and D.F.A.M.N., U. de Valencia,
     Avda. Dr. Moliner 50, ES-46100 Burjassot (Valencia), Spain
    \label{VALENCIA}}
\titlefoot{Institut f\"ur Hochenergiephysik, \"Osterr. Akad.
     d. Wissensch., Nikolsdorfergasse 18, AT-1050 Vienna, Austria
    \label{VIENNA}}
\titlefoot{Inst. Nuclear Studies and University of Warsaw, Ul.
     Hoza 69, PL-00681 Warsaw, Poland
    \label{WARSZAWA}}
\titlefoot{Fachbereich Physik, University of Wuppertal, Postfach
     100 127, DE-42097 Wuppertal, Germany \\
\noindent
{$^\dagger$~deceased}
    \label{WUPPERTAL}}
\addtolength{\textheight}{-10mm}
\addtolength{\footskip}{5mm}
\clearpage
\headsep 30.0pt
\end{titlepage}
%
\pagenumbering{arabic} 
\setcounter{footnote}{0} %
\large
%

\section{\label{introduction}Introduction}

The branching fractions of the $b$-quark into the different species of $b$-hadrons
are an important input and source of systematic uncertainty 
for many measurements in the heavy flavour sector where $b$-hadrons are
produced in jets,
e.g. analyses on $B$-meson oscillations or CKM elements at LEP.
Furthermore, these production fractions give insight into the fragmentation process.
Since $b$-quarks at LEP are mainly produced directly in the decay of the $Z^0$ boson, with
negligible contributions from later processes like gluon splitting $g\to\bb$,
$b$-hadron production fractions are sensitive to a certain step in the fragmentation
process, namely the beginning of the fragmentation chain.
This is not the case for analyses investigating inclusive particle production rates
of hadrons which do not contain a primary heavy quark.

The $b$-hadron production fractions are defined as the probability of a
$\overline{b}$- or $b$-quark to fragment into the corresponding $b$-hadron:
$\fbu = BR(\overline{b} \to \bplus)                  = BR(b \to B^-)$,
$\fbd = BR(\overline{b} \to \bnull)                  = BR(b \to \bnullbar)$,
$\fbs = BR(\overline{b} \to \bnulls)                 = BR(b \to \bnullsbar)$,
$\fbb = BR(\overline{b} \to \mbox{anti--$b$-baryon}) = BR(b \to \mbox{$b$-baryon})$.
Furthermore, the production fractions for charged and neutral $b$-hadrons
are defined as
$\fplus = BR(\overline{b} \to X^+_B)                 = BR(b \to X^-_B)$
and
$\fnull = BR(\overline{b} \to X^0_B)                 = BR(b \to X^0_B)$,
where $X^+_B$, $X^-_B$ and $X^0_B$ stand for any
positively charged, negatively charged or neutral
$b$-hadron, respectively.
With these definitions, $f_{B_i}$ is also the production fraction of
the $b$-hadron type $B_i$, particle or antiparticle, in $\bb$-events.

A direct measurement of these production fractions using exclusive decays is difficult,
since there are many decay channels with small branching fractions having
large relative uncertainties \cite{pdg02}.
For the determination of $\fbs$,
the inputs used are measurements of the product branching ratio
$BR(\overline{{b}}\to {B}_s^0)\cdot BR({B}^0_s\to {D}_s^-{l}^+\nu {X})$
at LEP \cite{dslana}, measurements of the ratio $\fbs/(\fbu+\fbd)$ using events with
exclusively reconstructed charm particles in semileptonic $b$-decays or
double semileptonic decays from CDF \cite{cdffractions}
and
the mixing parameters $\overline{\chi}$ and $\chi_d$.
The integrated mixing probability $\overline{\chi}$, in an unbiased sample of neutral
$B$-mesons, has contributions from $\bnull$- and $\bnulls$-mesons\footnote{Since $\overline{\chi}$
  is mainly measured using leptons, the rates $\fbd$ and $\fbs$ have to be weighted
  by ratios of lifetimes, $\tau_{B_d}/\tau_B$ and $\tau_{B_s}/\tau_B$, respectively.
  This has been omitted in the formulae for simplicity.}:
$\overline{\chi}=\fbd\chi_d+\fbs\chi_s$, where $\chi_d$ and $\chi_s$ are the integrated
mixing probabilities for $\bnull$- and $\bnulls$-mesons.
This allows the extraction of $\fbs$ quite precisely.
The baryon rate is estimated from similar product branching ratios,
using $\Lambda_c^+l^-$ and $\Xi^-l^-$ correlations \cite{cdffractions,adelbaryon},
and a measurement of proton production in $b$-hadron decays \cite{alephproton}.
No direct measurements of $\fbd$ or $\fbu$ have been published so far. 
The averages for weakly-decaying $b$-hadrons are
listed in \cite{pdg02}.
The following assumptions are made: the $b$-hadron production fractions are the
same in $Z\to\bb$ decays at LEP and in high-$p_t$ jets at the TEVATRON,
$\fbs + \fbb + \fbu + \fbd = 1$ and $\fbu=\fbd$.
The latter two are applied as constraints in the averaging procedure.
The combined result is
$\fbs=(10.6\pm 1.3)\%$,
$\fbb=(11.8\pm 2.0)\%$ and
$\fbd=\fbu=(38.8\pm 1.3)\%$.

\begin{figure}[tbh]
\begin{center}
\leavevmode 
\includegraphics[bb=0 0 1859 1297,width=\textwidth]{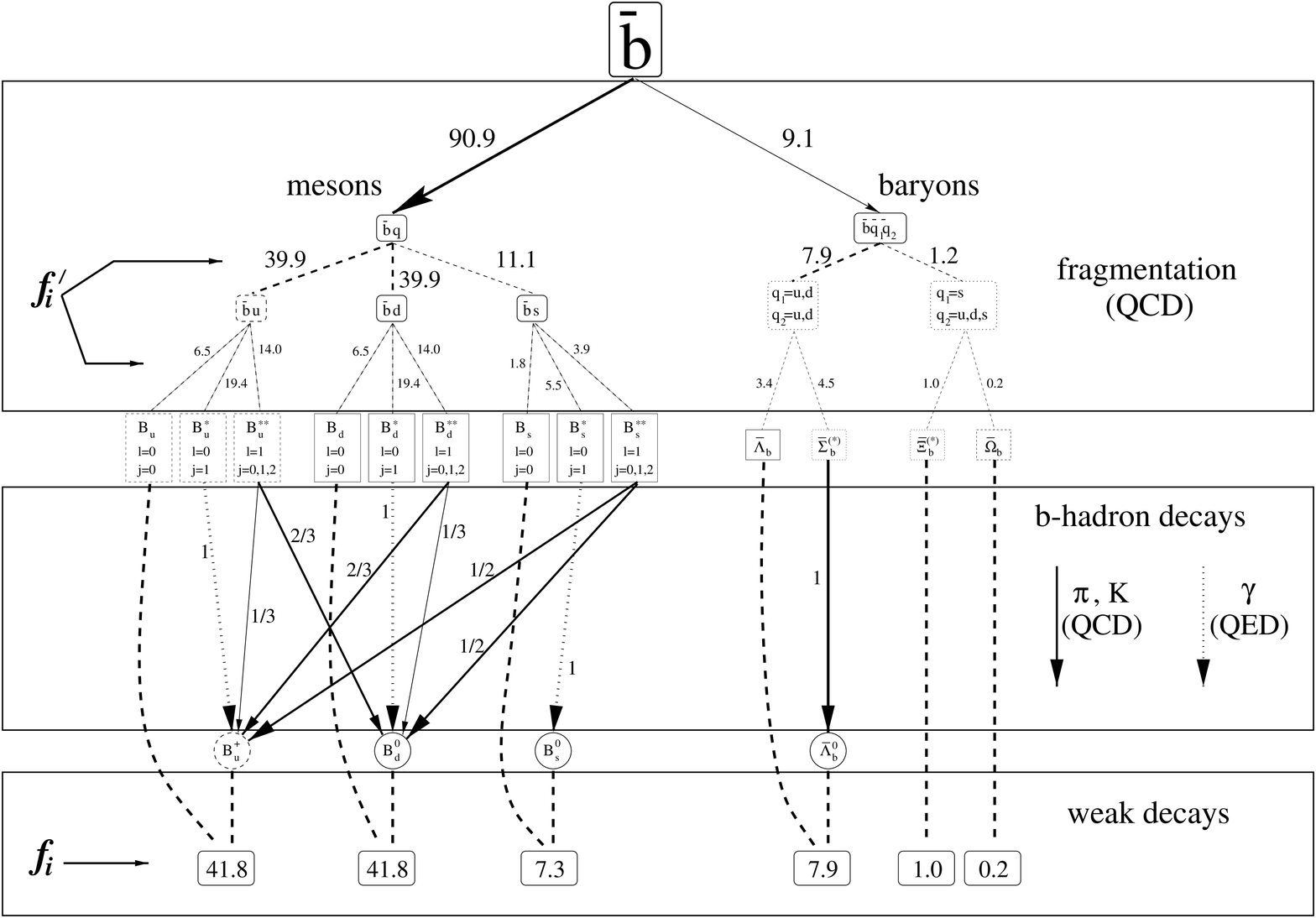}
\caption{\label{allrates} Schematic picture of the production mechanism of
                          $b$-hadrons. The rates, given in percent,
                          of hadrons primarily produced
                          in the fragmentation are denoted by $f'$, the ones which
                          decay through weak interaction
                          (indicated by dashed lines) by $f$.
                          Strong and electromagnetic decays are indicated by solid
                          and dotted arrows, respectively, together with
                          their (expected) branching ratios
                          (for strong decays only single pion and kaon
                           transitions have been considered).
                          The given rates are taken from simulation
                          (\jst Monte-Carlo model with parton shower option and
                           parameter settings
                           according to the DELPHI-Tuning \cite{deltune}).
                          The parameters giving the suppression of $\sq$ pairs and diquarks
                          are, respectively,
                          $\gamma_s=P(\sq)/P(\uu)=P(\sq)/P(\dd)$=0.28 and
                          $P(qq)/P(q)$=0.1.}
\end{center}
\end{figure}
In Figure \ref{allrates} a schematic picture of the $b$-hadron production process
is shown.
One has to distinguish between the fractions of $b$-hadrons which are directly
produced in the non-perturbative fragmentation process, denoted $\fbiprime$
in the following, and the fractions of weakly-decaying $b$-hadrons, $\fbi$.
Strong decays of
primarily
produced $b$-hadrons can lead to
$\fbiprime\ne\fbi$, e.g. the presence of orbitally excited
$B^{**}_s$-mesons with their expected decays
$B^{**}_s\to B^{(*)}K$ have the consequence $\fbsprime > \fbs$,
$\fbuprime < \fbu$ and $\fbdprime < \fbd$.
However, the equality $\fbuprime = \fbdprime$, which originates from isospin symmetry,
remains valid for weakly-decaying $b$-hadrons ($\fbu = \fbd$).\footnote{In contrast
  to the $D$-system, the presence of $B^*$-mesons does not change the rates of
  charged and neutral $B$-mesons, because their dominant decay mode is
  $B^*\to B\gamma$. This is also the case for orbitally excited
  $B^{**}$-mesons, if $f_{B_d^{**}}=f_{B_u^{**}}$, and isospin rules are
  used to calculate the dominant single pion transitions.}

For the rates of charged and neutral $b$-hadrons, an efficient algorithm has been
developed to distinguish charged particles from weak $B$-decays, from their fragmentation
counterparts produced at the primary event vertex.
This allows an estimate of the charge of the weakly-decaying
hadron to be made and thus a measurement of $\fplus$ and $\fnull$.
$\fbu$ can then be extracted from $\fplus$ with small additional uncertainties. 
The data taken in the years 1994 and 1995, when the DELPHI detector was equipped
with a double sided silicon vertex detector, have been used for the analysis.

The simulation used
the \jst model \cite{jetset} with parton shower option and parameters
determined from earlier QCD studies \cite{deltune},
followed by a detailed detector simulation \cite{delsim}.

\section{\label{experiment}The DELPHI detector and event selection}

The DELPHI detector is described in detail in references
\cite{detector,performance}.
The present analysis relies mainly on
charged particles, measured using information provided by
the central tracking detectors.
\begin{itemize}
\item The {\bf microVertex Detector} (VD) consists of three layers of silicon strip
      detectors at radii of 6.3, 9.0 and 10.9 cm. $R\phi$
      coordinates\footnote{In the standard DELPHI coordinate system, the $z$ axis is
        along the electron direction, the $x$ axis points towards the centre of LEP, and
        the $y$ axis points upwards. The polar angle to the $z$ axis is denoted by $\theta$,
        and the azimuthal angle around the $z$ axis by $\phi$; the radial coordinate is
        $R=\sqrt{x^2+y^2}$.}
      in the plane perpendicular to the beam
      are measured in all three layers. The first and third layers also
      provide $z$ information.
      The polar angle ($\theta$) coverage for a
      particle passing all three layers is from 44\degr to 136\degr.
      The single point precision has been estimated from real data to be
      about 8 $\mu$m in $R\phi$ and (for charged particles crossing
      perpendicular to the module) about 9 $\mu$m in $z$.
\item The {\bf Inner Detector} (ID) consists of an inner drift chamber with
      jet chamber geometry and 5 cylindrical
      layers of multiwire proportional chambers (MWPC).
      The jet chamber,
      between 12 and 23 cm in $R$ and 23\degr and 157\degr in
      $\theta$, consists of 24 azimuthal sectors, each providing up to
      24 $R\phi$ points. From 1995 on, a longer ID has been operational,
      with polar angle coverage from 15\degr to 165\degr and replacing the MWPC
      by 5 layers of straw tube detectors.
      The precision on local track elements has been measured in muon pair events
      to be about 45 $\mu$m in $R\phi$.
\item The {\bf Time Projection Chamber} (TPC) is the main tracking device of
      DELPHI. It provides up to 16 space points per particle trajectory
      for radii between 40 and 110 cm and polar angles between 39\degr and 141\degr.
      The precision on the track
      elements is about 150 $\mu$m in $R\phi$ and about 600 $\mu$m in $z$.
      For particle identification a measurement of the specific energy loss 
      (dE/dx) is provided by 192 sense wires located at the end caps of the
      drift volume.
\item The {\bf Outer Detector} (OD) consists of 5 layers of drift tubes
      between radii of 197 and 206 cm. Its polar angle coverage is from
      42\degr to 138\degr. The OD provides 3 space points and 2 $R\phi$
      points per track.
      The single point precision is about 110 $\mu$m in the
      $R\phi$ plane and about 3.5 cm in the $z$ direction.
\end{itemize}
An event has been selected as multihadronic if the following
requirements are satisfied:
\begin{itemize}
\item There must be at least 5 charged particles in the event, each with
      momentum larger than 400 MeV/$c$ and polar angle between 20\degr
      and 160\degr.
\item The total reconstructed energy of these
      charged particles has to exceed 12\% of the centre-of-mass energy
      (assuming all particles to have the pion mass).
\item The total energy of the charged particles in each hemisphere
      (defined by the plane perpendicular to the beam axis)
      has to exceed 3\% of the centre-of-mass energy.
\end{itemize}

After these cuts, about 2 million events from the 1994 and 1995 runs
have been retained.
About 7 million simulated
${Z}^0 \to {q} \overline{{q}}$ and
3.1 million ${Z}^0 \to {b} \overline{{b}}$ events
have been selected with the same cuts.

Jets have been reconstructed with the LUCLUS algorithm \cite{jetset} ($d_{join}=5$ GeV/$c$)
using charged and neutral particles.
Two-jet events well within the acceptance of the vertex
detector ($|\cos\theta_{thrust}|<0.65$) were selected.
The event was divided in two hemispheres by the plane perpendicular to the
thrust axis.

The most important variables to tag or antitag $\bb$ events are
the track impact parameters of charged particles with respect to the primary vertex
which is fit on an event-by-event basis using the position and size of the
beamspot as constraints.
From the track impact parameters and their errors a probability is computed
that a selected sample of charged particles originates from the primary vertex.
To increase efficiency and purity, additional information
e.g. from reconstructed secondary vertices and identified leptons
are used. A combined discriminating variable is then used to select $\bb$ events. 
These methods are described in detail in \cite{btaggingpaper}.
The tagging of $\bb$ events was performed in the hemisphere opposite to the one
which was used for the measurement.
The cut on the discriminating variable of the combined $b$-tagging has been chosen 
to give a $b$-purity of about 97.5\%.
A secondary vertex was fit in the hemisphere considered for the measurement.
Hadronic interactions in the detector material were reconstructed using the
algorithm described in \cite{mammoth}. Since the particle causing the interaction is lost
in most of the cases, hemispheres with such interactions were rejected.

\section{\label{bcharge}Measurement of the rates of charged and neutral $b$-hadrons}

\begin{figure}[bth]
\begin{center}
\leavevmode 
\includegraphics[bb=27 2 846 418,width=\textwidth]{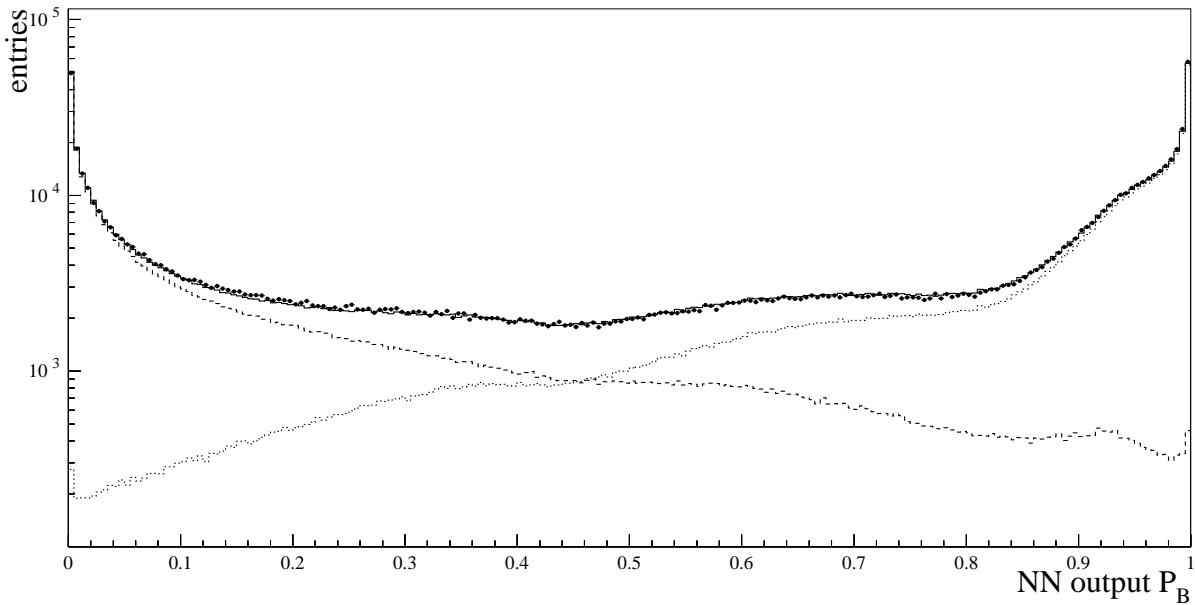}
\caption{\label{netoutput} The output of the neural network used to separate $B$-decay particles
                           from fragmentation particles.
                           Shown are the data (closed circles),
                           the simulation (solid histogram) and the contributions from 
                           weak $B$-decay particles (dotted) and their
                           fragmentation counterparts produced at the primary vertex (dashed).
                           The latter distribution includes particles originating from
                           strong decays of excited $b$-hadrons.}
\end{center}
\end{figure}
The basic idea for measuring the rates of charged and neutral $b$-hadrons
is to reconstruct the charge of the weakly-decaying hadron.
Based on a neural network, for each charged particle in a hemisphere, a probability
$P_B$ that the particle
originates from a $b$-hadron decay rather
than from fragmentation is calculated.
Charged particles are accepted if their momentum exceeds 500 MeV/$c$ and if at least one
vertex detector hit has been associated.
At least four charged particles had to be accepted in the hemisphere.
The maximum number of charged particles in the hemisphere failing these
acceptance cuts was limited to four.
The input variables to the neural network are the probability
that the charged particle track fits to the primary vertex,
the momentum, the rapidity of the particle with respect to the thrust axis,
the reconstructed flight distance
from the primary to the secondary vertex
in the $R\phi$ plane and its error.
The last two are not specific for the particles
but for the hemisphere considered. They give additional information about the
separation power of the other variables, especially the vertex probability.
The distributions of the net output variable\footnote{This variable can be
interpreted as Bayesian a-posteriori probability. $P_B$ is computed from this
variable taking into account the ratio of $b$-hadron decay particles and fragmentation
particles as taken from simulation. This is necessary because the neural network has been
trained with equal numbers of charged particles from these two classes.} 
are shown in Figure \ref{netoutput}.\footnote{If not explicitely
stated otherwise, all figures show the data from 1994 and 1995.}

A secondary-vertex charge $Q_B$ is then constructed through
\begin{equation}
  \label{eqqb}
 Q_B = \sum_{i=1}^{N_{hem}} Q_iP_{B,i} \quad,
\end{equation}
where $N_{hem}$ is the number of accepted particles in the hemisphere,
$Q_i$ the charge of particle $i$ and $P_{B,i}$ its probability to stem from
a $b$-hadron decay as defined above.
Assuming binomial statistics, an error on $Q_B$ can be defined as
\begin{equation}
  \label{errorqb}
 \sigma_{Q_B} = \sqrt{\sum_{i=1}^{N_{hem}} P_{B,i}(1-P_{B,i})}\quad .
\end{equation}
This quantity does not account for particle losses due to inefficiencies
in the track reconstruction. $\sigma_{Q_B}$ is small if all charged particles are well
classified, having values of $P_B$ close to 0 or 1, and gets larger the
more particles have probabilities around 0.5.

Parameters in the simulation possibly having an effect on the measurement
have been adjusted to their measured values. They are discussed in detail
in section \ref{systsection} and listed also in Table \ref{errorsyst}.

\begin{figure}[bt]
\begin{center}
\leavevmode 
\includegraphics[bb=8 5 450 415,width=0.48\textwidth]{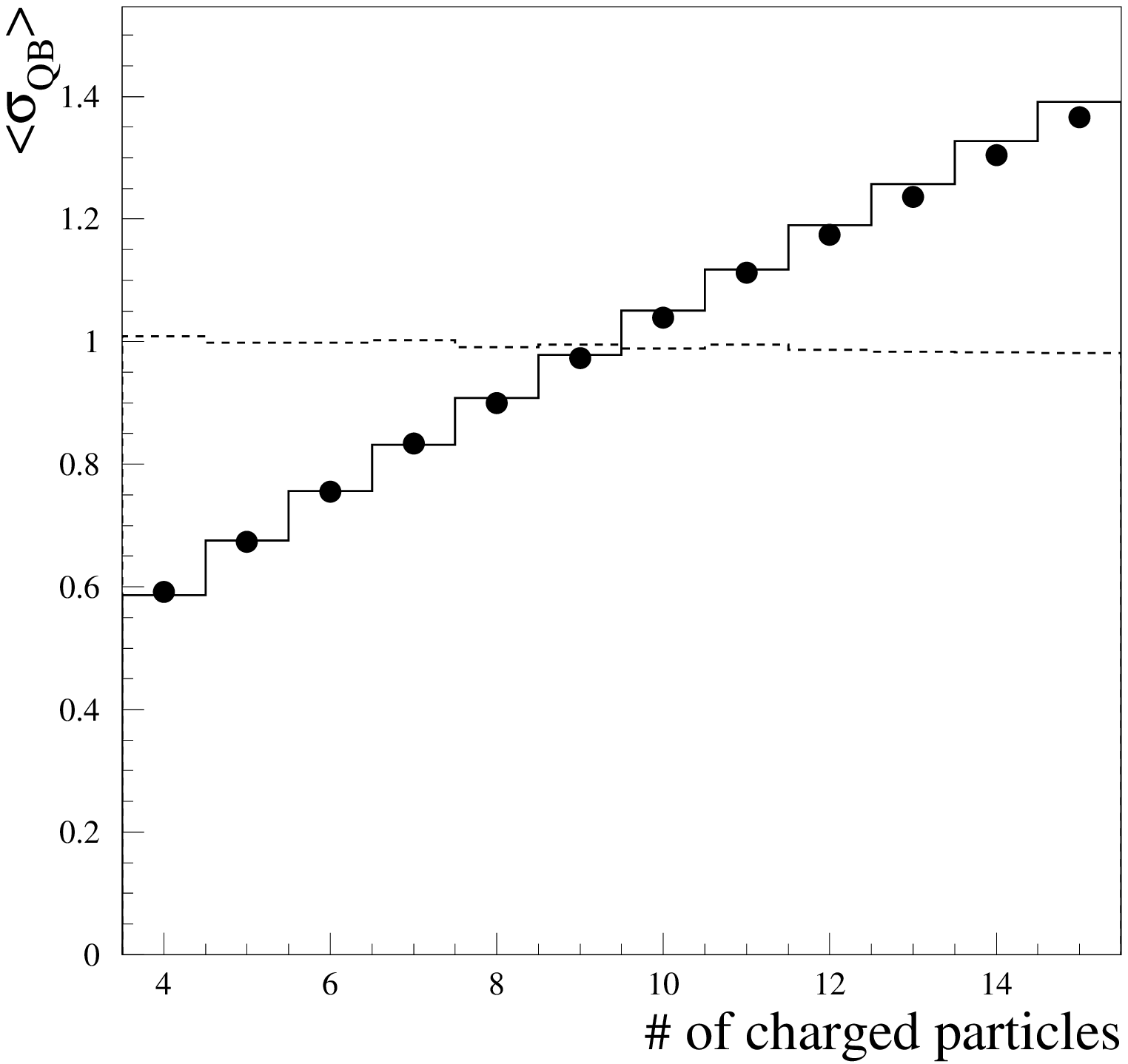}
\includegraphics[bb=1 5 450 415,width=0.48\textwidth]{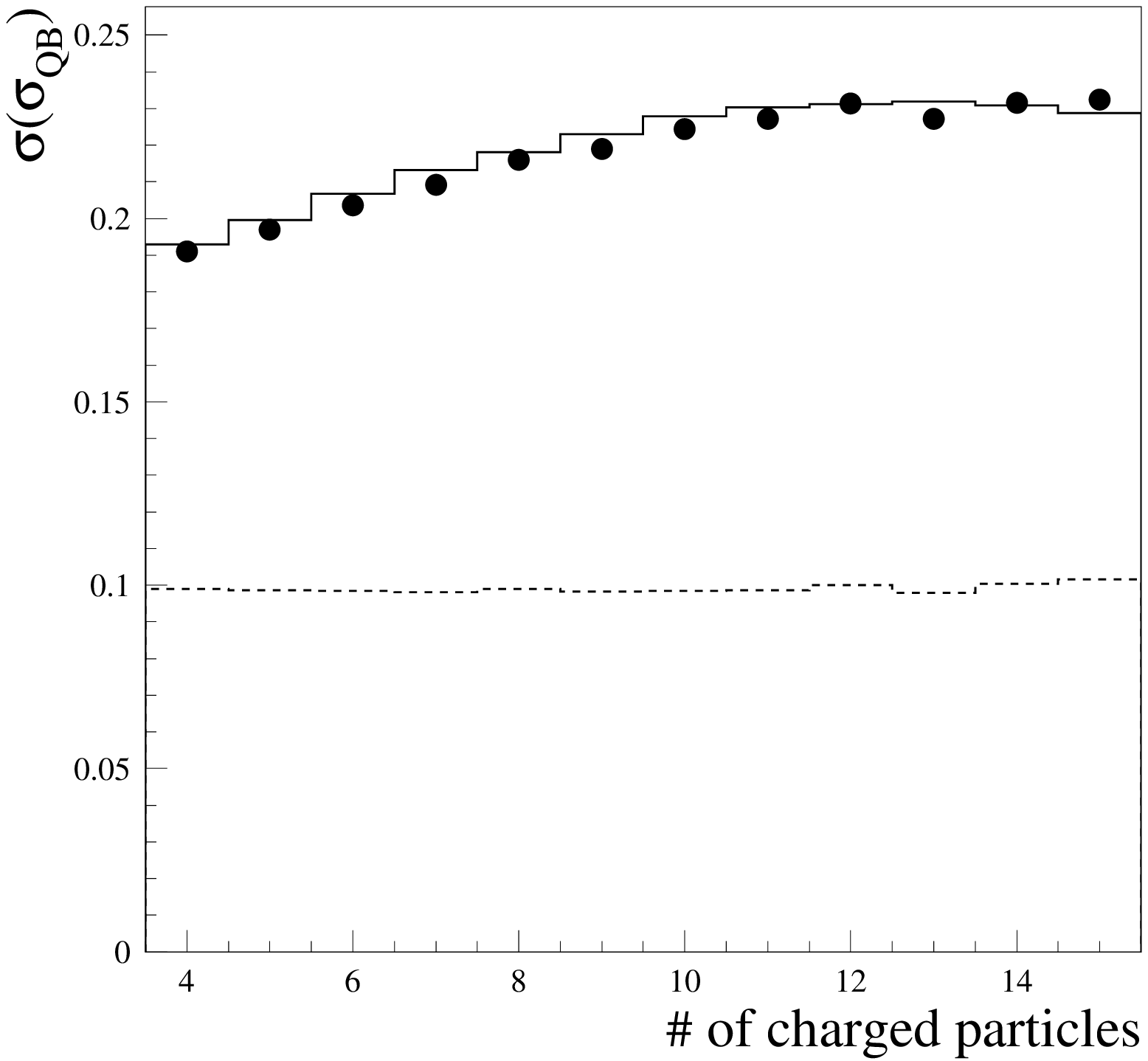}
\caption[hihi]{\label{evertvsntr} $\langle\sigma_{Q_B}\rangle$ (left) and
                            $\sigma(\sigma_{Q_B})$ (right) versus the number
                            of charged particles which
                            have been used
                            for the estimation of the vertex charge for
                            data (closed circles) and simulation (histogram)
                            before applying the correction as explained in the text.
                            The ratios
                            $\langle\sigma_{Q_B}\rangle_{data}/\langle\sigma_{Q_B}\rangle_{sim.}$
                            and
                            $0.1\cdot \sigma(\sigma_{Q_B})_{data}/\sigma(\sigma_{Q_B})_{sim.}$
                            are shown as dashed lines. The deviations are within $\pm 3\%$ over
                            the whole range.}
\end{center}
\end{figure}
The shapes of the $Q_B$ distributions are directly affected by the number of
charged particles which have
been used in the reconstruction of the vertex charge. This number is larger in the data
than in the simulation by 0.12 at a mean value of about 8.0.
The shape of the particle multiplicity distribution is very similar in the data and the simulation.
The simulated events are reweighted to get agreement in this distribution.
The error of the vertex charge, defined in equation \ref{errorqb}, also influences
the $Q_B$ distributions because it is directly related to its width.
Incorrect modelling of the shape of the $Q_B$ distribution
in the simulation potentially biases the result.
The dependence of the mean value and spread of $\sigma_{Q_B}$
on the number of charged particles is shown in Figure \ref{evertvsntr}.
Data and simulation agree well, the deviations being within $\pm 3\%$.
The simulated events are
reweighted to get these
distributions in agreement, individually for any number
of charged particles, to avoid any possible bias. 
To correct for the loss of charged particle tracks,
the number of charged particles in the hemisphere not
passing the track cuts is also brought into agreement.
All these corrections are small.
Comparing the $Q_B$ distributions of data and simulation shows slight shifts
of the data distribution with respect to the simulation
which is corrected depending on the
absolute value of the polar angle $\theta$
(these shifts are typically around 0.01 with the deviations from zero not
being very significant).
Such shifts can be caused by small differences in the material distribution
of the detector in data and simulation causing a charge asymmetry.

\begin{figure}[tbh]
\begin{center}
\leavevmode 
\includegraphics[bb=12 12 425 330,width=0.45\textwidth]{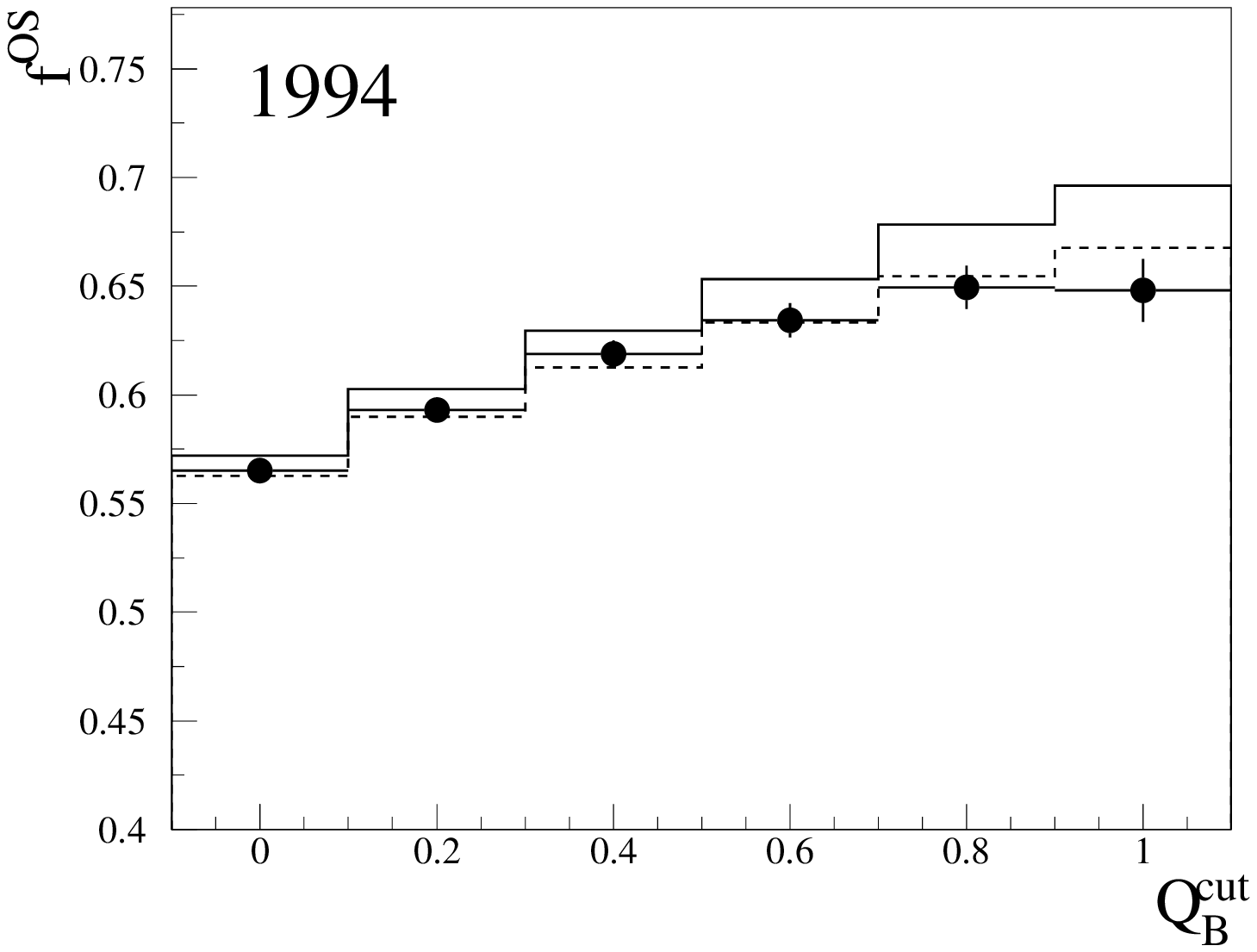}
\includegraphics[bb=12 12 425 330,width=0.45\textwidth]{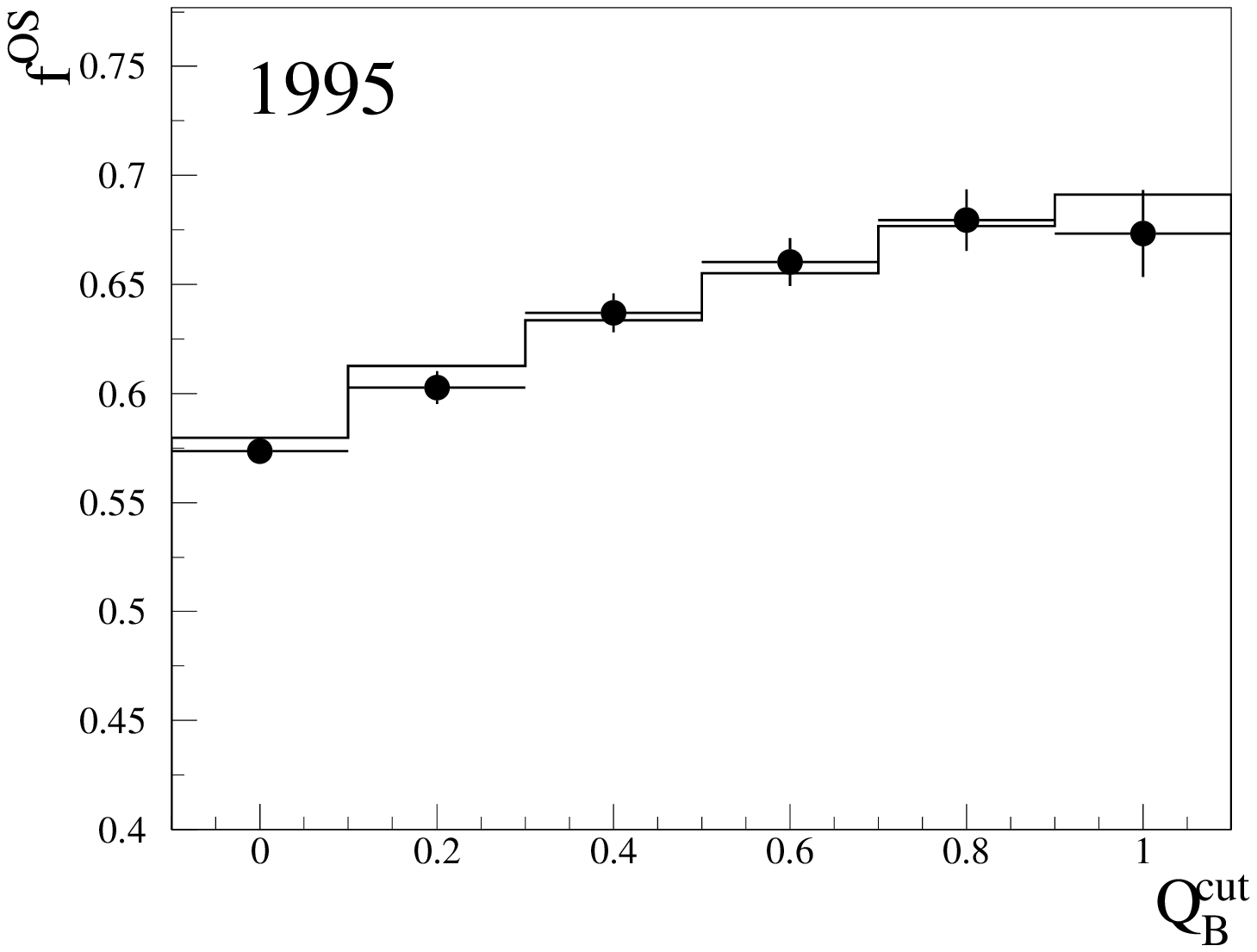}
\caption{\label{qcalibfig} The fraction of `opposite-sign' events versus $Q_B^{cut}$
                           for 1994 (left) and
                           1995 (right) comparing data (circles with error bars) and
                           simulation (histogram), used for the calibration of $Q_B$.
                           For 1994, the solid (dashed) histogram shows the simulation
                           before (after) applying the correction explained in the text.}
\end{center}
\end{figure}
The vertex charge $Q_B$ is sensitive to the charge of the $b$-quark in the hemisphere,
mainly in cases where charged $b$-hadrons are produced.
Since $b$- and $\bbar$-quarks are produced in pairs in $Z$ decays, 
events where the vertex charge $Q_B$ can be determined in both hemispheres
(`double tagged' events in the following)
can be used to calibrate $Q_B$ on the data themselves.
This is done in the following way.
For 20116 events, where all cuts are passed in both hemispheres,
`opposite-sign' (called `$OS$' in the following) and
`same-sign' (`$SS$') events are defined through the vertex charges
in both hemispheres:
$OS$: $Q_B^1 \cdot Q_B^2 < 0$,
$SS$: $Q_B^1 \cdot Q_B^2 > 0$.
To be sensitive to the shape of the distributions,
$|Q_B^{1,2}| > Q_B^{cut}$ has been required.
The fractions of $OS$ and $SS$ events are directly related to the probability to
tag the charge of the $b$-quark in the hemisphere correctly.
In Figure \ref{qcalibfig}, the fraction of $OS$ events, $f^{OS}$, versus $Q_B^{cut}$ is shown
for the 1994 and 1995 data sets. It can be seen that the probability to
tag the $b$-quark charge correctly is slightly overestimated in the simulation for 1994
(with a tendency to become worse when increasing $Q_B^{cut}$, thus probing
the wings of the distribution), whereas
good agreement is found for 1995.
The fraction of $OS$ events is mainly determined by the $Q_B$ distributions
of charged $b$-hadrons. 
To correct for the disagreement between data and simulation in 1994,
the $Q_B$ distributions for positively and negatively
charged $b$-hadrons are reweighted in the simulation
according to
$w^{\pm} = 1 - (\pm a_1 \cdot Q_B - a_2)^3$.
The following
parameters were found to make the fraction of $OS$ events
in the simulation consistent with the data over almost the full range of $Q_B^{cut}$
(see Figure \ref{qcalibfig}):
$a_1=0.31 \mbox{\ and\ } a_2=0.34$.
Another, related, distribution is the fraction of double tagged events
versus $Q_B^{cut}$ with respect to all double tagged events.
It has been verified that this distribution is in good agreement
for both years of data taking (after applying the correction to the 1994 simulation).

\begin{figure}[tbh]
\begin{center}
\leavevmode 
\includegraphics[bb=6 7 896 827,width=0.9\textwidth]{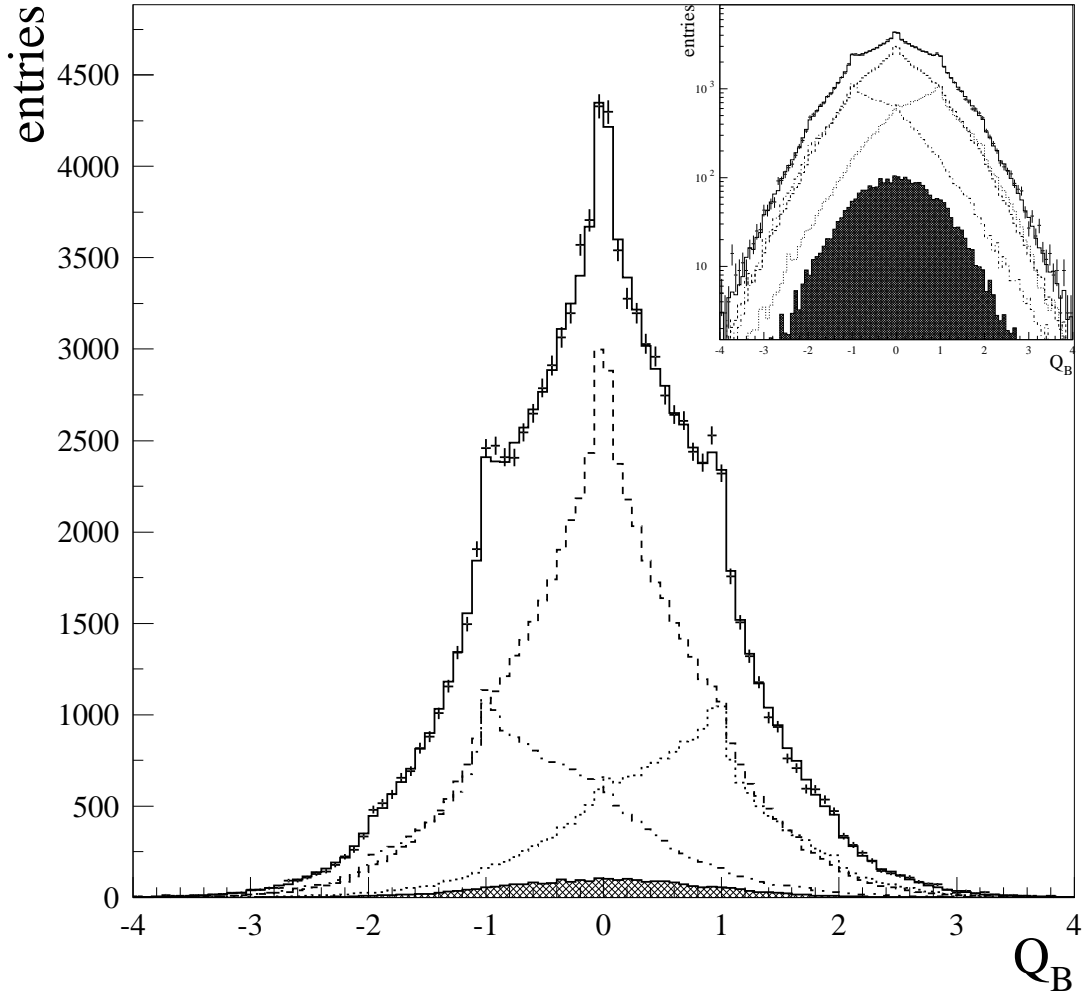}
\caption{\label{qfit} Distribution of the vertex charge, $Q_B$, of the weakly-decaying b-hadron
                      for the data (points with
                      error bars) with the result of the fit
                      superimposed (solid histogram) on a linear scale and
                      on a logarithmic scale (as inlay).
                      The shapes for neutral (dashed histogram),
                      negatively (dashed-dotted) and positively (dotted) charged
                      $b$-hadrons obtained from the simulation
                      (in the fit, one single component was used for positively and negatively
                       charged $b$-hadrons)
                      are also shown.
                      The hatched histogram shows the contribution of non-$\bb$ events.}
\end{center}
\end{figure}
The measured $Q_B$ distribution has been fit by the corresponding shapes expected for charged
and neutral $b$-hadrons obtained from the simulation,
while not separating the shapes for the positive and negative charges.

A technique based on a binned log-likelihood method taking into account
the limited statistics of the simulation has been used \cite{fitlimmc}.
The non-$\bb$ background
has been fixed to the value obtained from simulation.
The real data distribution, corresponding to 103.285 selected hemispheres,
together with the fit result and the simulation prediction for neutral,
positively and negatively charged $b$-hadrons is shown in Figure \ref{qfit}.
The result for
$\fplus$, the fraction of charged $b$-hadrons in a sample of
weakly-decaying $b$-hadrons produced in $Z^0\to\bb$ decays,
is
$\fplus = ( 41.84 \pm 0.99 \mbox{(stat.)} ) \%$ for the 1994 data set and
$\fplus = ( 42.65 \pm 1.48 \mbox{(stat.)} ) \%$ for 1995
giving a combined value of
\begin{equation}
  \fplus = ( 42.09 \pm 0.82 \mbox{(stat.)} ) \%.
\end{equation}
The result for
$\fnull$ is given through
$\fnull = 1-\fplus = ( 57.91 \pm 0.82 \mbox{(stat.)} )\%$,
with $\fplus$ and $\fnull$ being fully anticorrelated. 
The $\chi^2$ per degree of freedom
of the fits are 0.96 and 1.06 for the two years.

\section{\label{systsection}Systematic checks and uncertainties}

Several cross-checks have been performed.
The fit range
and number of bins
of the histograms used in the fit have been varied.
The momentum cut has been varied in the range from 300 to 800 MeV/$c$
and the maximum number of rejected tracks in the hemisphere between two and five.
No significant change of the result has been found.
The distributions of negatively and positively charged $b$-hadrons have been
fit separately. This gives
\begin{eqnarray*}
  BR(b,\overline{b}\to X^+_B) & = & ( 20.66 \pm 0.60 \mbox{(stat.)} ) \% \\
  BR(b,\overline{b}\to X^-_B) & = & ( 21.16 \pm 0.59 \mbox{(stat.)} ) \%.
\end{eqnarray*}
for 1994 and
\begin{eqnarray*}
  BR(b,\overline{b}\to X^+_B) & = & ( 21.37 \pm 0.87 \mbox{(stat.)} ) \% \\
  BR(b,\overline{b}\to X^-_B) & = & ( 21.29 \pm 0.86 \mbox{(stat.)} ) \%.
\end{eqnarray*}
for 1995.
The two numbers are correlated
($\rho^\pm \approx 0.44$).
\begin{figure}[tbh]
\begin{center}
\leavevmode 
\includegraphics[bb=0 8 510 385,width=0.7\textwidth]{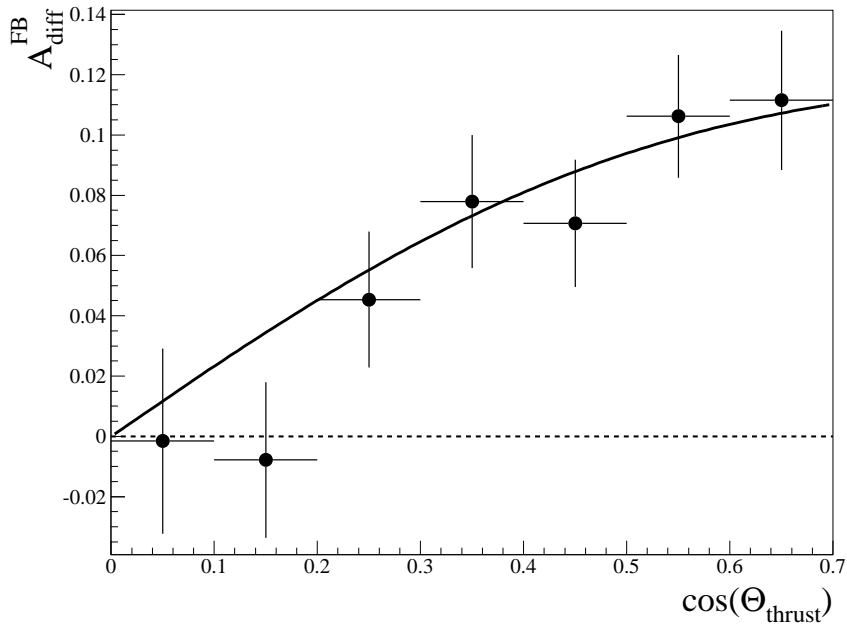}
\caption[]{\label{qvertvsthrust}The dependence of the
                                measured differential asymmetry $A^{FB}_{diff}$ on
                                $\cos\theta_{thrust}$ for the data.
                                $A^{FB}_{diff}$ has been computed from the vertex charges in
                                the forward and backward hemispheres as:
                                $A^{FB}_{diff}=
                                 (<Q_B^{FW}>-<Q_B^{BW}>) / <Q_B^b>$,
                                where $<Q_B^b>$ is the mean of the vertex charge for hemispheres
                                containing a $b$-quark.
				The errors indicated are statistical only.
                                The expectation for the pole asymmetry $A^{FB}_b=0.0982$
                                (from \cite{pdg02})
				is superimposed as solid line.}
\end{center}
\end{figure}
As already mentioned and used for the calibration,
the vertex charge $Q_B$ can be used as flavour tag for hemispheres where the $b$- or
$\overline{{b}}$-quark fragmented into a charged $b$-hadron and is thus sensitive
to the forward-backward asymmetry $A^{FB}_b$.
The differential asymmetry, computed from the vertex charges in the forward and
backward hemispheres, versus the direction of the thrust axis is shown in
Figure \ref{qvertvsthrust}, showing good agreement with the expectation from
the measured value of the pole asymmetry $A^{FB}_b=0.0982\pm 0.0017$ (from \cite{pdg02}).

\begin{table}[tbh]
\begin{center}
\begin{tabular}{|c|c|r|}\hline
source & value and variation & $\delta\fplus[\%]$ \\ \hline\hline
$\tau_{B^0}$             & $(1.542 \pm 0.016)$ ps  & $+0.007$ \\
$\tau_{B^+}$             & $(1.674 \pm 0.018)$ ps  & $-0.129$ \\
$\tau_{B_s}$             & $(1.461 \pm 0.057)$ ps  & $+0.027$ \\
$\tau_{b-baryon}$        & $(1.208 \pm 0.051)$ ps  & $+0.010$ \\
$\chi_d$                 & $0.181 \pm 0.004$       & $+0.038$ \\
$\fbs$                   & $( 8.5 \pm 1.3)$ \%     & $+0.217$ \\
$\fbb$                   & $( 9.5 \pm 2.0)$ \%     & $+0.057$ \\
$f_{\Xi_b^-}$            & $( 1.1 \pm 0.5)$ \%     & $-0.187$ \\
$P_{B^{**}}$             & $0.24 \pm 0.04$         & $+0.287$ \\
$P_{\Sigma_B^{(*)}}$     & $0.10 \pm 0.05$         & $-0.027$ \\
wrong sign charm rate    & $(20.0 \pm 3.3)$ \%     & $-0.001$ \\
$BR(B^{0(+)}\to \overline{D}^{-(0)}+X)$& text      & $-0.183$ \\
b fragmentation function & text                    & $ 0.201$ \\
%
min. \# of acc. charged particles   & 4,3          &  $ 0.217$ \\
$Q_B$ calibration        & text                    &  $ 0.677$  \\
non-$\bb$ background     & $\pm 30\%$               & $+0.113$ \\ \hline \hline
total                    &                         & $ 0.886$ \\ \hline
\end{tabular}
\end{center}
\caption{\label{errorsyst}
         Breakdown of systematic errors on $\fplus$.
         For the $b$-hadron fractions, $\fbs$ and $\fbb$,
         their correlation has been
         taken into account, for the other sources of systematic uncertainties the
         errors have been added in quadrature.
         The signs of the errors given are for an upwards variation of
         the corresponding physics parameter.
         For $BR(B^{0(+)}\to \overline{D}^{-(0)}+X)$, an upwards variation means 
         increasing $BR(\bnull\to\dminus+X)$ and adjusting the other, related
         branching ratios as explained in the text.}
\end{table} 
To estimate systematic errors, the following parameters in the simulation and
cuts have been varied:
\begin{itemize}
  \item Lifetimes of $b$-hadrons.
  \item The oscillation frequency and thus the mixing probability $\chi_d$ for $\bnull$-mesons.
        Mixing of $\bnull$-mesons leaves the contribution from
        fragmentation particles unchanged but reverses the flavour of the weakly-decaying
        $B$-meson. 
        The contributions to the vertex charge from $\bnull$- and $\bnullbar$-mesons
        are slightly different. One reason is that $\bnull$-mesons produce
        dominantly $D^-$-mesons whereas $\bnullbar$-mesons mainly give $D^+$, leading
        to different charges at the tertiary charm vertex.
  \item The rates of different $b$-hadron species,
        because different $b$-hadrons
        contribute to a
        certain charge (e.g. $\bnull$, $\bnulls$, $\lambdab$, $\Xi_b^0$ to $Q=0$) and
        their distributions look slightly different.
        The value of $\fbu(=\fbd)$ has been set to the measured value in this analysis.
        The values of $\fbs$ and $\fbb$ from \cite{pdg02} have been rescaled accordingly to
        ensure $\fbs + \fbb + \fbu + \fbd = 1$.
  \item The rates of excited $b$-hadrons, namely orbitally excited $B^{**}_{u,d,s}$-mesons
        and $\Sigma_b^{(*)}$-baryons. The expected strong decays of these states produce
        particles (pions or kaons) looking partly like fragmentation particles
        (coming from the primary vertex) and partly like $B$-decay particles (having
         large rapidity).
        From the results in \cite{bdoublestar,delphispecichep02}
        an average
        $P_{B^{**}_{u,d}}=
         BR(\overline{{b}}\to\bdstarud)/BR(\overline{{b}}\to \overline{{b}}({u,d}))
         =0.24\pm 0.04$
        is computed.
        It is assumed that $P_{B^{**}_s}=P_{B^{**}_{u,d}}$.
        For $\Sigma_b^{(*)}$ production DELPHI gives an upper limit in 
        \cite{delphispecichep02}. The following
        value is taken as a conservative choice\footnote{$P_{\Sigma_b^{(*)}}$ is
        defined in the same way as $P_{B^{**}}$.}: $P_{\Sigma_b^{(*)}}=0.10\pm 0.05$.
  \item The rate of so called `wrong-sign charm' production at the upper $W$ vertex from
        \cite{hfsteer01}.
  \item The branching ratios of `right-sign' decays of $\bplus$- and $\bnull$-mesons
        into $\dbarnull$- and $\dminus$-mesons.
        Because of $\tau({\dminus})>\tau({\dbarnull})$, $b$-hadron decays with a
        $\dminus$ in the final state have charged particles from the tertiary charm decay which
        are more displaced from the primary vertex than in the case of a $\dbarnull$,
        affecting the vertex variables used as input to the neural network.
        The following branching ratios have been used, being consistent with the
        measured inclusive production rates of $\dbarnull$- and $\dminus$-mesons in
        $B$-decays:
           $BR(\bnull\to\dminus+X)  =15.6\%$,
           $BR(\bnull\to\dbarnull+X)=65.8\%$,
           $BR(\bplus\to\dminus+X)  =29.3\%$,
           $BR(\bplus\to\dbarnull+X)=52.1\%$.
        These branching ratios have been varied by 5\% (absolute).
        The variation has been performed in a correlated way to keep the total rate
        of $\dbarnull$- and $\dminus$-mesons in $B$-decays and
        $BR(\bnull/\bplus \to \dbarnull+X) + BR(\bnull/\bplus \to \dminus+X)$ constant. 
  \item The $b$-quark fragmentation function.
        The measured
	function from
	DELPHI \cite{delphibfrag}
	has been used.
        For the systematic error the full difference to the model implemented
        in the simulation (Peterson with $\epsilon_b=0.002326$) has been taken.
        This is a conservative choice, since the measurement errors are much smaller
        than the deviations from the fragmentation function in the simulation.
  \item The cut on the minimum number of accepted charged particles in the hemisphere used for
        the calculation of $Q_B$ has been changed from an even to an odd number.
        This could have a systematic effect because charged (neutral) $b$-hadrons decay to an
        odd (even) number of charged particles.
  \item The calibration of the vertex charge (described in Section \ref{bcharge}).
        The parameters
	$a_1$ and $a_2$
	have been varied independently in a way
	which results in a displacement of the $f^{OS}$ versus $Q_B^{cut}$ curve
	(Figure \ref{qcalibfig}) in the simulation,  
	corresponding to a shift by one standard deviation with respect to the data errors.
	This procedure has been consistently applied also to the
	1995 year simulation, even if no correction has been applied in this case,
	because of the good agreement between data and simulation.
	The error from the parameter giving the largest variation in $\fplus$ has
	been chosen.
  \item The non-$\bb$ background (mainly $\cc$) has been varied by $\pm 30\%$.  
\end{itemize}
If not explicitly stated otherwise, the corresponding numbers have been
taken from \cite{pdg02}.
The breakdown of the systematic errors
is shown in Table \ref{errorsyst}.
The result, including systematic errors, is:
\begin{equation}
  \fplus = ( 42.09 \pm 0.82 \mbox{(stat.)} \pm 0.89 \mbox{(syst.)} ) \%.
\end{equation}

\section{\label{interpretation}Interpretation of the results}

Weakly-decaying neutral $b$-hadrons are $\bnull$- and $\bnulls$-mesons, the $\lambdab$-baryon
and the strange $b$-baryon $\Xi_b^0$.
The rate of charged $b$-hadrons
is dominated by the $\bplus$-meson, strange $b$-baryons giving only a minor
contribution ($\Xi_b^-$, $\Omega_b^-$). Formally, one gets:
\begin{eqnarray}
 \label{ratescontributions}
 \fnull &=& \fbd + \fbs + \fbbnull, \nonumber \\
 \fplus &=& \fbu + \fbbplus.
\end{eqnarray}
In \cite{hfsteer01}, the fraction of $\Xi_b^-$-baryons has been estimated analysing
production rates of $\Xi^-l^-$ final states:
$f_{\Xi_b^-} = BR(\overline{b}\to\overline{\Xi}_b^+)  = BR(b\to\Xi_b^-)  = (1.1\pm 0.5)\%$.
This fraction is subtracted from $\fplus$ to get the fraction of
$\bplus$-mesons in a sample of weakly-decaying $b$-hadrons
produced in the fragmentation of $\bbar$-quarks 
(the production fraction of $\Omega_b^-$-baryons is expected to be negligible).
The result is
\begin{equation}
  \fbu = (40.99 \pm 0.82 \mbox{(stat.)} \pm 1.11 \mbox{(syst.)})\%,
\end{equation}
where the error from $f_{\Xi_b^-}$ has been added to the systematic error
from $\fplus$ taking into account the correlation arising from the fact that
the uncertainty of $f_{\Xi_b^-}$ is a source of uncertainty for $\fplus$
(compare Table \ref{errorsyst}).

\section{\label{conclusions}Conclusions}

A precise measurement of the production fractions of weakly-decaying charged
and neutral $b$-hadrons has been presented for the first time.
The fraction of $\bbar$-quarks fragmenting into positively charged
weakly-decaying $b$-hadrons,
and thus the fraction of charged $b$-hadrons in a sample of weakly-decaying
$b$-hadrons produced in $Z^0\to\bb$ decays,
has been measured to be
$\fplus = ( 42.09 \pm 0.82 \mbox{(stat.)} \pm 0.89 \mbox{(syst.)} ) \%
= ( 42.09 \pm 1.21 ) \% $.
Subtracting the $\overline{\Xi}_b^+$
rate (assuming that the $\overline{\Omega}_b^+$ rate is negligible) gives
$\fbu = (40.99 \pm 0.82 \mbox{(stat.)} \pm 1.11 \mbox{(syst.)})\% = (40.99 \pm 1.38)\%$.
This is, so far, the most precise dedicated measurement of a production fraction of a
specific $b$-hadron.
The accuracy on $\fbu$ is comparable to the accuracy achieved by combining all other
information available on $b$-hadron production fractions. 
This measurement thus represents significant new information that will impact
on combinations of measurements 
to estimate $b$-hadron production fractions in jets.

\subsection*{Acknowledgements}
\vskip 3 mm
 We are greatly indebted to our technical 
collaborators, to the members of the CERN-SL Division for the excellent 
performance of the LEP collider, and to the funding agencies for their
support in building and operating the DELPHI detector.\\
We acknowledge in particular the support of \\
Austrian Federal Ministry of Education, Science and Culture,
GZ 616.364/2-III/2a/98, \\
FNRS--FWO, Flanders Institute to encourage scientific and technological 
research in the industry (IWT), Federal Office for Scientific, Technical
and Cultural affairs (OSTC), Belgium,  \\
FINEP, CNPq, CAPES, FUJB and FAPERJ, Brazil, \\
Czech Ministry of Industry and Trade, GA CR 202/99/1362,\\
Commission of the European Communities (DG XII), \\
Direction des Sciences de la Mati$\grave{\mbox{\rm e}}$re, CEA, France, \\
Bundesministerium f$\ddot{\mbox{\rm u}}$r Bildung, Wissenschaft, Forschung 
und Technologie, Germany,\\
General Secretariat for Research and Technology, Greece, \\
National Science Foundation (NWO) and Foundation for Research on Matter (FOM),
The Netherlands, \\
Norwegian Research Council,  \\
State Committee for Scientific Research, Poland, SPUB-M/CERN/PO3/DZ296/2000,
SPUB-M/CERN/PO3/DZ297/2000 and 2P03B 104 19 and 2P03B 69 23(2002-2004)\\
JNICT--Junta Nacional de Investiga\c{c}\~{a}o Cient\'{\i}fica 
e Tecnol$\acute{\mbox{\rm o}}$gica, Portugal, \\
Vedecka grantova agentura MS SR, Slovakia, Nr. 95/5195/134, \\
Ministry of Science and Technology of the Republic of Slovenia, \\
CICYT, Spain, AEN99-0950 and AEN99-0761,  \\
The Swedish Natural Science Research Council,      \\
Particle Physics and Astronomy Research Council, UK, \\
Department of Energy, USA, DE-FG02-01ER41155, \\
EEC RTN contract HPRN-CT-00292-2002. \\

\clearpage


\end{document}